\documentstyle[epsf]{l-aa}

\newcommand{\aaa}{A\&A }      
\newcommand{\apj}{ApJ }       
\newcommand{\mnras}{MNRAS }   
\newcommand{\bm}[1]{\mbox{{\boldmath $#1$}}}

\newcommand{\fr}[2]{\frac{\displaystyle #1}{\displaystyle #2}}

\newcommand{\pder}[3]{\fr{{\partial}^{#3} {#1}}{{\partial} {#2}^{#3}}}

\newcommand{\km}{{\rm \ km}}
\newcommand{\cm}{{\rm \ cm}}
\newcommand{\pc}{{\rm \ pc}}
\newcommand{\kpc}{{\rm \ kpc}}
\newcommand{\s}{{\rm s}}

\newcommand{\Myr}{{\rm \ Myr}}
\newcommand{\dyn}{{\rm \ dyn}}

\renewcommand{\Im}{\rm Im}
\begin{document}

\thesaurus{02(02.13.1; 02.09.1; 11.13.2; 11.19.2; 11.09.4; 09.11.1)}

\title{The galactic dynamo effect due to Parker-shearing instability of
magnetic flux tubes}
\subtitle{II. Numerical simulations and the nonlinear evolution.}

\author{M. Hanasz}

\institute{Centre for Astronomy, Nicolaus Copernicus University,
 PL-87-148 Piwnice/Torun, Poland, ({\em mhanasz@astri.uni.torun.pl})
}

\offprints{M. Hanasz}

\date{Received November 26, 1996/ accepted May 29, 1997}

\maketitle

\markboth{M.  Hanasz: The galactic dynamo effect due to Parker-shearing
instability of magnetic flux tubes }{}

\begin{abstract}
In this paper we continue investigations of the Parker-shearing instability
performing numerical simulations of the magnetic flux tube dynamics in the thin
flux-tube approximation.  We show that evolution of flux tubes resulting
from numerical simulations is very similar to that of linear solutions if the
vertical displacements are smaller than the vertical scale height $H$ of the
galactic disc.  If the vertical displacements are comparable to $H$, the
vertical growth of perturbations is faster in the nonlinear range than in the
linear one and we observe a rapid inflation of the flux tube at its top,
which leads to a singularity in numerical simulations, if only the cosmic rays
are taken into account.  Then we perform simulations for the case of nonuniform
external medium, which show that the dominating wavelength of the Parker
instability is the same as the wavelength of modulations of external medium.
As a consequence of this fact, in the case of dominating cosmic ray pressure,
the dynamo $\alpha$ effect related to these short wavelength modulations is
much more efficient than that related to the linearly most unstable long
wavelengths modes of the Parker instability.  Under the influence of
differential forces resulting from differential rotation and the density waves,
the $\alpha$-effect is essentially magnified in the spiral arms and
diminished in the interarm regions, what confirms our previous
results obtained in the linear approximation.


\end{abstract}

\section{Introduction}

In the preceding paper (Hanasz and Lesch  1997, hereafter Paper
I) we investigated
the idea that the galactic dynamo action is driven by the Parker-shearing
instability (Parker 1992, Hanasz \& Lesch 1993,
hereafter HL'93).  We use the terms ``Parker instability'' or ``Parker-shearing
instability'' depending on the context (i.e.  on the current relevance of the
rotational shearing effects).  We presented the theory of slender magnetic flux
tubes, analogous to that proposed by Spruit (1981), Spruit \& Van Ballegooijen
(1982) and Moreno-Insertis (1986), adopted to the case of galactic magnetic
flux tubes.  Then, performing the linear stability analysis we calculated the
dynamo coefficients resulting from the Parker-shearing instability of magnetic
flux tubes in the thin flux tube approximation.  The calculations allowed us to
derive a series of results which made conspicuous the power of the
Parker-shearing instability as the mechanism generating the dynamo action
(see Sect.  4.  of Paper I).

It is obvious, however, that the linear approximation does not allow to inquire
those aspect of the problem which are related to the evolution of perturbations
for large amplitudes compared to the vertical scale height $H$.  The
problem results from the fact that the perturbations grow exponentially in the
linear approximation and after a finite time period some physical quantities
reach unphysical values.  Since the possibilities of the extension of the work
in the frame of analytical methods are rather limited, we decided to continue
our studies by means of numerical simulations.  We are going to present
numerical solutions of the set of nonlinear equations describing the dynamics
of flux tubes derived in Sect.  2 of Paper I together with the related
dynamo transport coefficients.  The solutions will allow us to verify
the results of Paper I in the nonlinear regime and make a point to an old
question about the final state of the Parker instability in galactic discs.

The problem of the final state of the Parker instability with and without cosmic
rays has been discussed intensively in the past.  Parker (1968, 1969) argues
that cosmic rays leave the Galaxy by rapidly inflating bubbles on the magnetic
field lines.  This picture has been denied by Mouschovias (1974, 1975)
who has shown that: 1) without cosmic rays the Parker instability attains
final, curved equilibrium states and 2) the final equilibria are attained also
in the presence of cosmic rays if only the number of cosmic ray particles
within a particular flux tube is conserved during the flux tube motion.  In
this model the cosmic ray pressure is uniform along the magnetic field line and
varies with the volume of the flux tube.  These 2-dimensional solutions of
Mouschovias have been tested with respect to instability against the
perturbations in the 3-rd (radial) dimension by Asseo et al.  (1978, 1980),
Lechieze-Rey et al.  (1980) and appeared to be unstable even without cosmic
rays.

Some approaches to the Parker instability treated the cosmic ray gas as a
relativistic gas, whose particles propagate with infinite speed along the
magnetic field lines (see the discussion by Shu (1974) on this topic and
references therein).  It appears however, that the cosmic ray gas streaming
along the magnetic field lines induces Alfven waves which exert a friction and
slow the effective transport of cosmic rays down to the Alfven speed.  Parker
(1992) discusses his concept of inflating magnetic lobes, assuming explicitly
that the inflation rate is limited by the Alfvenic streaming spend of the
cosmic ray gas.

The assumption of Mouschovias on the conservation of the number of cosmic ray
particles within the flux tube cannot be retained because of the high
production rate of cosmic rays due to the supernova explosions.  There are at
least two observational facts which strongly support the Parkers point of view.
1) The measured cosmic ray lifetime $10^7$ yr in galactic discs implies that
the cosmic rays should disconnect from the disc together with the trapping
magnetic field.  The characteristic time scale of Parker instability $\sim
10^7$ yr is consistent with this point of view.  2) In some edge-on galaxies a
vertical magnetic field is observed in radio continuum (Hummel et al.  1991;
Golla and Hummel 1994) extending in galactic halos.  One should mention that
the Parker instability is invoked to explain the formation of vertical magnetic
fields in the galactic halo, among the other models like galactic
fountains (Kahn and Brett, 1993) and galactic winds (Spencer and Cram, 1992).
We are going to point out however, that the Parker instability with the
contribution of cosmic rays can serve as a fully sufficient explanation.

In HL'93 and Paper I we assumed that flux tubes are anchored at heavy molecular
clouds.  We have proposed that the typical intercloud distance determines the
characteristic wavelength of the Parker instability.  From Fig.5 of Paper I
it follows that the range of unstable wavelengths depends on the
magnetic pressure (via $\alpha$). In the limit of weak magnetic field
(e.g. $\alpha=0.001$) the marginal stability point is far below the wavelength
100 pc, then the typical distance between clouds ($\sim 100 \pc$) corresponds
to an unstable wavelength.  Analogously, we can expect that even with the
present magnetic field strength an instantaneous and local growth of the cosmic
ray pressure due to supernova explosions can help to overcome the magnetic
field tension and shift the marginal stability point below given intercloud
distance.  In such a case the dominating wavelength of Parker instability can
be much smaller than the frequently invoked value of a few vertical
scaleheights.

In the present section we generalize the physical setup and introduce a
physically new effect - the propagation of flux tubes in an external medium
which is nonuniform.  The nonuniformities of the external medium are assumed
to be spatially periodic and oscillating in time.  It is our aim to check what
is the evolution of a single flux tube in the nonuniform environment.  The
local and temporal external gas condensations are intended to imitate the
clouds in the ambient medium.  We assume that such clouds appear and disappear
along the length of the flux tube modifying the external density only, but have
not a dynamical influence on the flux tube.  This seems to be quite difficult
at present to treat consistently the flux tube -- cloud, or flux tube -- flux
tube collisions.  Such a local and temporal external density modulations
influence the flux tube internal mass distribution via the magneto-hydrostatic
balance condition (Eqn.  (30) of Paper I).  Variations of the external density
distribution force a continuous mass redistribution inside the tube triggering
the possible instabilities all over the flux tube lifetime, not only at the
initial moment.  In this respect our model seems to be more similar to
of realistic interstellar medium than to the highly idealized uniform case.  We
are going to show that even a small amplitude external density modulations
determine the characteristic wavelengths of the Parker modes.  This in turn
essentially influence the magnitude of the $\alpha$-effect.

The effects of the galactic dynamics on the Parker instability has been
discussed in Paper I in the linear approximation.  The differential forces due
to the axisymmetric differential rotation and the density waves influence
essentially the dynamics of flux tubes as well as the dynamo transport
coefficients.  We found in Paper I that due to the presence of the radial
differential force the curves of growth rate split into two separate parts at
the wavenumber $k_{crit}$ which is determined by the balance of the
differential force, Coriolis force and the radial component of magnetic
tension.  In the range $k_{crit} < k < k_{marg}$ (the Parker range), where
$k_{marg}$ is the ordinary marginal stability point, the linear solutions are
stationary growing, as in the case without shear and the $\alpha$-effect is
essentially magnified.  In the range $0 < k < k_{crit}$ (the shearing range)
the linear solutions behave like propagating waves and the $\alpha$-effect is
essentially diminished.  We pointed out that the density waves can modify the
differential force periodically with a relatively large amplitude, so that a
perturbation with a given wavenumber placed in vicinity of $k_{crit}$ can move
from the Parker to shearing range and vice versa depending on the phase of the
density wave.  Since the Parker range corresponds to the spiral arms and the
shearing range to the interarm regions we showed that the $\alpha$-effect can
switch on in arms and -off in the interarm regions.  We predict that the
regular component should be stronger in the interarm regions than in the arms
due to the simultaneous action of strong shear
and the magnetic reconnection.  These properties are very consistent with the
observed arms of the uniform magnetic field in NGC 6946 (Beck and Hoernes 1996)
placed in between the optical arms.  We are going to verify the above results
in the nonlinear range by means of the nonlinear numerical simulations.

The plan of this paper is as follows:  In Section 2.  we present the applied
numerical method, Section 3 is devoted to the comparison of the numerical
solutions with linear analytical solutions of Paper I.  In Section 4 we
analyze the final states of the Parker instability.  In Section 5 we calculate
the temporal evolution of the dynamo coefficients.  In Section 6 the physical
conditions are generalized by introducing modulations in the density of
external medium, in which the flux tube propagate.  In Section 7 we examine the
influence of the effects of differential forces in the nonlinear regime on
the flux tube evolution and the dynamo action.  Conclusions and possible
relations of our theoretical results to the observational results are contained
in Section 8.

\section{The method}

In Paper I we have derived the system of MHD equations governing evolution of a
slender magnetic flux tube in the thin flux tube approximation.  Geometry of
the thin flux tube and the equations of motion have been described in the
Section 2.  of Paper I.  The system of partial differential equations
describing evolution of the flux tube is given by equations (35), (36), (37) of
Paper I for 9 unknown functions of time and position along the flux tube,
together with the condition of magnetohydrostatic equilibrium of the flux tube
with its surrounding (34).  The unknown functions are:  the Lagrangian
coordinates of an element of the flux tube $x$, $y$, $z$, the components of the
tangent vector $L_x$, $L_y$, $L_z$ and the components of velocity $v_x$, $v_y$
and $v_z$.  We shall not rewrite these equations here for brevity.  Since we
are going to solve the system of partial differential equations by means of 1D
numerical simulations we shall concentrate only on aspects related directly to
this method.

We shall use the numerical algorithm PDECOL taken from the NETLIB
library (algorithm 540R, Madsen \& Sincovec 1992).  The package PDECOL is based
on the method of lines and uses a finite element collocation procedure (with
piecewise polynomials as the trial space) for the discretization of the spatial
variable $x$.  The collocation procedure reduces the partial differential
equation system to a semi-discrete system which then depends only on the time
variable $t$.  The time integration is then accomplished by use of standard
techniques.  The package solves the general system of nonlinear partial
differential equations of at most second order of the form

\begin{equation}
\pder{U}{t}{} = F\left(t,x,U,\pder{U}{x}{},\pder{U}{x}{2}\right)
\end{equation}
with given boundary conditions and an initial condition.  In our approach we
shall apply the periodic boundary conditions and the initial conditions of two
types.  The first type will be given by an eigenmode
resulting from the linear stability analysis.  Such an initial state will allow
to compare the linear analytical and the nonlinear numerical solutions.
Results obtained with this initial state will serve for tests of the numerical
algorithm.  As the second type of the initial state we shall use a random
superposition of periodic perturbations of the initial velocity and the
vanishing initial displacements.  Initial states of this type will allow for
studies of the flux tube dynamics in conditions which more realistically
resemble the nonuniform interstellar medium.

\section{The comparison of analytical and numerical solutions}

With the aim to test the correctness of the numerical solutions we shall compare them
with the analytical linear solutions of the Paper I.  First, we shall apply the
initial state of a small amplitude, which is given by the linear solution with
a wavenumber $k$ and an amplitude $Z_{init}$.  Then we shall observe the
temporal
evolution of the numerical and analytical solutions which both start from the
same initial state.  We expect that before the nonlinear effects become
important, these two solutions should evolve accordingly.  We are going to
compare the linear and the nonlinear amplitudes of displacements of the flux
tube gas in the three spatial directions at a given moment $t_0$.  The linear
solution will be represented by $X_l = |X_1| \exp (\omega_i t_0) $, $Y_l =
|Y_1| \exp (\omega_i t_0) $ and $Z_l = |Z_1| \exp (\omega_i t_0) $.  The
numerical nonlinear solution will be represented by the appropriate quantities
$X_n = \max|\Delta X(s,t_0)|$, $Y_n=\max|\Delta Y(s,t_0)|$ and $Z_n =
\max|\Delta Z(s,t_0)| $, where $\Delta X(s,t_0)$, $\Delta Y(s,t_0)$ and $\Delta
Z(s,t_0) $ are the Lagrangian displacements of an element of the flux tube with
respect to its equilibrium position at the time $t_0$.  The maxima are taken
over a full range of variability of the length parameter $s$.  An exemplary
result obtained with the angular velocity typical for our Galaxy $\Omega =
\Omega_G$, the Oort constant $A=0$, the ratios of cosmic ray and magnetic
pressures to the gas pressure $\alpha=\beta=1$, the aerodynamic drag
coefficient $C_D=0$ and the wavelengths of the perturbation $H = 1330
\pc$ (representing the most unstable mode) is shown in Fig.~1.
\begin{figure}
\epsfxsize=\hsize \epsfbox{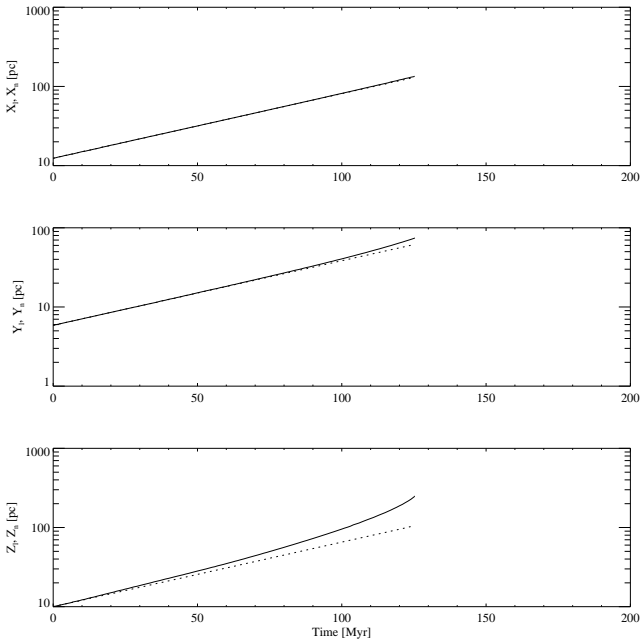}
\caption[]{
The comparison of the nonlinear numerical solution (full line) to the linear
analytical solution (dotted line). The amplitudes of displacements in three
directions vs. time are plotted. The basic parameters are:
$\Omega = \Omega_G$, $A=0$, $\alpha=\beta=1$, $\lambda=1330 \pc$ and $C_D = 0$
}
\end{figure}
\begin{figure}
\epsfxsize=\hsize \epsfbox{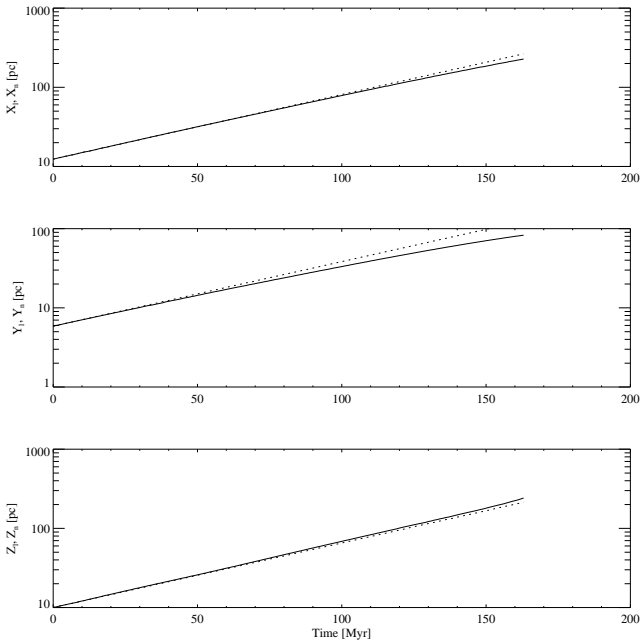}
\caption[]{
The comparison of the nonlinear numerical solution (full line) to the linear
analytical solution (dotted line). The amplitudes of displacements in three
dimensions vs. time are plotted. Parameters like in Fig.~1
but $C_D=1$ and $r=10 \pc$.}
\end{figure}
Throughout the
paper the initial height of the azimuthal flux tube is $Z_0 = 100 \pc$, which
is comparable to a half of the typical vertical scaleheight ($H\sim 200 \pc$).
The other parameters, which are not mentioned
explicitly are the same as those assumed in Paper I for our Galaxy.

We note that the solutions are in a very good agreement especially
concerning the coordinates $x$ and $y$.  However, a small discrepancy in the
vertical direction is remarkable.  The reason for this discrepancy is
in the linear approximation of expression for the magnetic tension.  Since in
the linear approximation the curvature vector $\bm{K}$ has only the linear
component, then the coefficient $B^2/4\pi\rho_i$ contributes only in the zeroth
order.  This means that we take in linear approximation $B_0^2$ instead of
$B^2$, which is overestimated at the upward displaced parts of the flux tube,
inflated due to the cosmic rays.  Then the vertical component of the
magnetic tension is diminished in the nonlinear regime with respect to the
linear approximation.  A similar argumentation applies also to the buoyancy
term.  This term is underestimated in the linear approximation.  This explains
why the top of the flux tube rises faster in numerical simulations than in the
linear approximation.  Thus, we can say that the numerical and analytical
linear solutions are mutually consistent for relatively small vertical
amplitudes $Z < H$.  This allows us to expect that the basic conclusions
of Paper I,  essential for the dynamo theory remain invariable in the nonlinear
regime for $Z < H$.  We shall demonstrate in the next sections of this
paper that this expectation is correct.  Moreover, we shall be able to take
into account some effects, which have not been tractable within the frame of
the linear approximation.

As an example we shall examine the effect of aerodynamic drag force.  HL'93
proposed that the aerodynamic drag force plays an important role
for the dynamo transport coefficients, especially for the turbulent
diffusivity.  It was argued in that paper that the aerodynamic drag force slows
down the vertical motions of fluxtubes and thereby diminishes the vertical
diffusion of magnetic field.  The discussion of the effect of the drag force
has been omitted in Paper I since we focused our attention on the linear
approximation and the resulting properties.  Solving the flux tube equations by
means of the nonlinear numerical simulations we are able to take the drag
force into account.  A corresponding solution with the parameter set as in
Fig.~1 but with the aerodynamic drag force determined by the aerodynamic drag
coefficient $C_D =1$ and the flux tube radius $r=10 \pc$, is presented in
Fig.~2.

It is apparent in Fig.~2 that with the given values of parameters, the
aerodynamic drag force influences motion of the flux tube only moderately.
One should be aware however that the assumed values of the drag coefficient and
the flux tube radius are essentially model dependent.  Moreover, the assumed
physics underlying the aerodynamic drag effect may appear to be too simple
for the case of galactic flux tubes.  Nevertheless, one can notice that
according to our expectations the drag force slows down the components of
velocity perpendicular to the flux tube axis retaining the longitudinal
velocity almost unchanged.

\section{The final state of Parker instability}
In this section we shall continue to examine the nonlinear evolution of
modes, which are introduced to the numerical experiment as the linear initial
states, as it was described in the previous section.  The first point, which
deserves a special attention, is related to the final states of our
simulations.  We note that the simulations brake down in a certain moment
$T_{end}$.  Moreover, we are not able to extend the simulations past $T_{end}$
by making the grid resolution more tiny.  Typically we divide the spatial
domain -- the full period of the perturbation into 200 grid cells.  Even the
division of the spatial domain into 5000 cells does not allow to extend
simulations past $T_{end}$.  The more frequent sampling of the numerical
solutions allows to approach $T_{end}$ closer and closer.  In the following
Figs.~3.  and 4.  we illustrate what happens.  We plot all the most relevant
physical variables:  3 components of the Lagrangian displacement of the flux
tube, together with the 3 components of velocity and acceleration, the internal
density $\rho_i$, the external density $\rho_e$, the strength of magnetic field
$B$ inside the tube, the mass per unit length $\xi$, the flux tube radius $r$
and the ratio of the total speed to the Alfven speed $|v|/v_A$ as a function of
position along the tube $s$.  Fig.~3.  and Fig.~4.  show these quantities at
$T= 125.0 \Myr$ and $T = 125.2 \Myr$, respectively
($T_{end} < 125.3 \Myr$ ).
\begin{figure}
\epsfxsize=\hsize \epsfbox{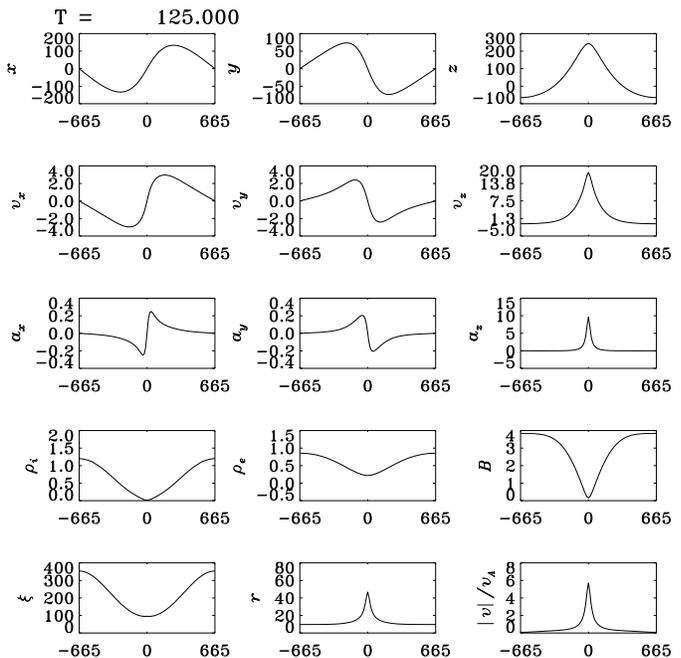}
\caption[]{The snapshot of the most relevant physical variables vs. position
along the flux tube  $s$, short before the brake down of the simulation.
The basic parameters are:
$\Omega = \Omega_G$, $A=0$, $\alpha=\beta=1$, $\lambda=1330 \pc$ and $C_D = 0$
The units are: pc, Myr, 1H-atom/cm$^3$ and $\mu G$.
The plotted quantities are:
the components of displacement, velocity and acceleration, internal and
external densities $\rho_i$, $\rho_e$, the internal magnetic field, mass per
unit length $\xi$, radius of the flux tube $r$ and the ratio of the modulus of
velocity to the Alfven speed $|v|/v_A$.  }
\end{figure}
\begin{figure}
\epsfxsize=\hsize \epsfbox{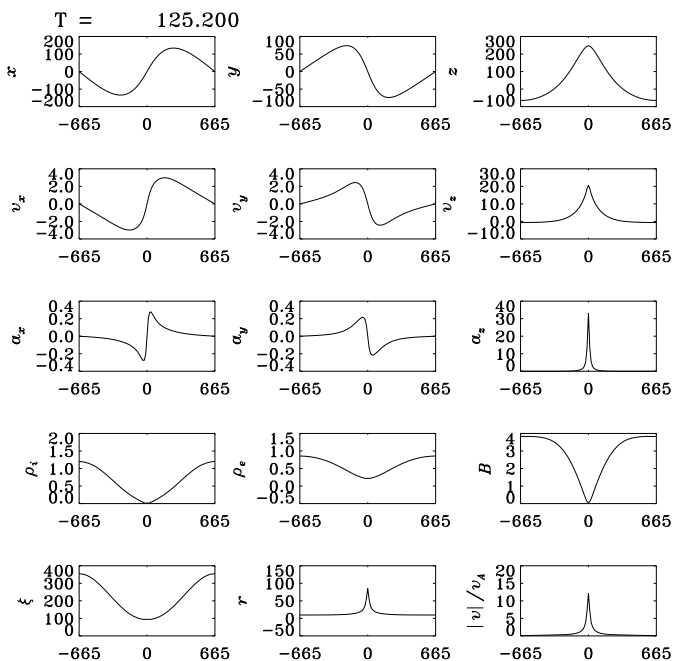}
\caption[]{The same as in previous figure 0.2 Myr later, but still before the
break down.}
\end{figure}

We can easily notice that the internal density $\rho_i$ approaches 0 at the top
of flux tube at the moment directly preceding $T_{end}$, while the external
density still remains nonvanishing.  The vertical acceleration due to the
buoyancy force starts to grow without limit within a limited time period.  The
appropriate panels of Figs.~3 and 4.  apparently show this effect.  All the
other quantities behave regularly up to the moment $T_{end}$.  It is worthwhile
however, to have a look at the other panels too.

First of all, we notice that the topmost part of the tube starts to inflate
(the flux tube radius $r$ grows) at the time preceding $T_{end}$.  We shall
argue that the reason for this inflation is the cosmic ray pressure.  We have
assumed: 1) that the cosmic ray pressure is constant along the magnetic field
lines and 2) that it is fixed at $z_0$ (as if the flux tube were connected to
an infinite reservoir of cosmic rays at $z_0$).
The gas density and pressure of the unperturbed disc decay
exponentially with the increasing height $z$, thus at a certain height $z_1$
the total (gas + magnetic + cosmic ray) pressure equals to the cosmic ray
pressure at the initial flux tube height $z_0$.  The cosmic ray pressure inside
the flux tube element displaced from $z_0$ to $z_1$ is the same as
the one at $z_0$.  Then, this is not possible to balance the internal cosmic
ray pressure if the flux tube element is  displaced further above
$z_1$, where the surrounding total pressure is smaller than the cosmic ray
pressure alone inside the raised element of the flux tube.  This pressure
imbalance has to result in a rapid inflation of the rising flux tube at its
top.  The inflation together with the outflow of the thermal gas from the
top cause that the internal gas density rapidly approaches
zero.  Since the buoyancy force is still nonvanishing and the internal mass
density becomes infinitely small, the vertical acceleration has to become
infinitely large.  We propose to call the effect described above as a {\em top
singularity} for brevity.

There are however weak points of the presented scenario. The magnetic flux is
conserved, the magnetic field diminishes together with the internal gas
density in such a way that the internal Alfven speed grows, but the vertical
acceleration grows even faster, so that the vertical speed of the top starts to
exceed the internal Alfven speed (see Figs.~3.  and 4., bottom-right panel).
As it was already mentioned (Hartquist and Morfil, 1986; Parker 1992) the bulk
streaming speed of the cosmic ray gas is limited to the Alfven speed
approximately, thus the assumption on the infinite streaming speed is not
fulfilled in the final stages of our simulations.  The limited streaming speed
is difficult to incorporate quantitatively in the actual considerations, but
one can expect that this effect should drastically limit the inflation rate as
soon as the top speed starts to exceed the Alfven speed (the cosmic rays are
not able to follow the top).  Finally, the top speed should saturate at the
value of the order of Alfven speed, which is large anyway.

On the other hand, the thin flux tube approximation becomes
invalid in the numerical simulations as the flux tube diameter becomes
comparable to the vertical scale lengths, what happens very close to the top
singularity.  The top singularity causes that we are not able to trace the
evolution as well as the properties relevant to the dynamo theory after the
moment $T_{end}$, even if all other parts of the flux tube in spite of its top
(and its surrounding) behave regularly.

Concluding this section we can say that the top singularity is an important
limitation for any numerical investigations of the Parker instability which
does not take into account the limited propagation speed of cosmic rays along
the magnetic field lines.
\section{The dynamo coefficients}

\begin{figure}
\epsfxsize=\hsize \epsfbox{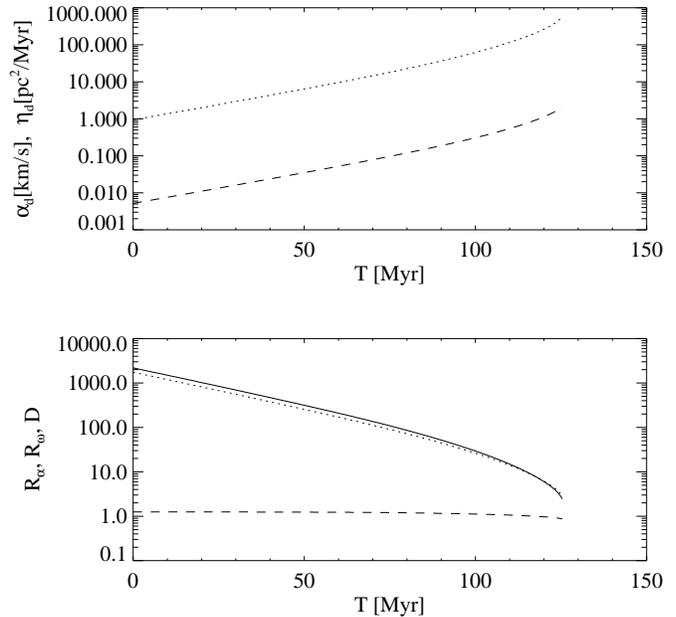}
\caption[]{
The dependence of coefficients $\alpha_d$ (dashed line) and $\eta_d$ (dotted
line) on time is presented in the upper panel.  The associated dynamo numbers
$R_{\alpha}$ (dashed line), $R_{\omega}$ (dotted line) and $D$ (full line) are
shown in the lower panel.  The aerodynamic drag coefficient $C_D=0$ and the
other parameters are like in Fig.~1.
}
\end{figure}
\begin{figure}
\epsfxsize=\hsize \epsfbox{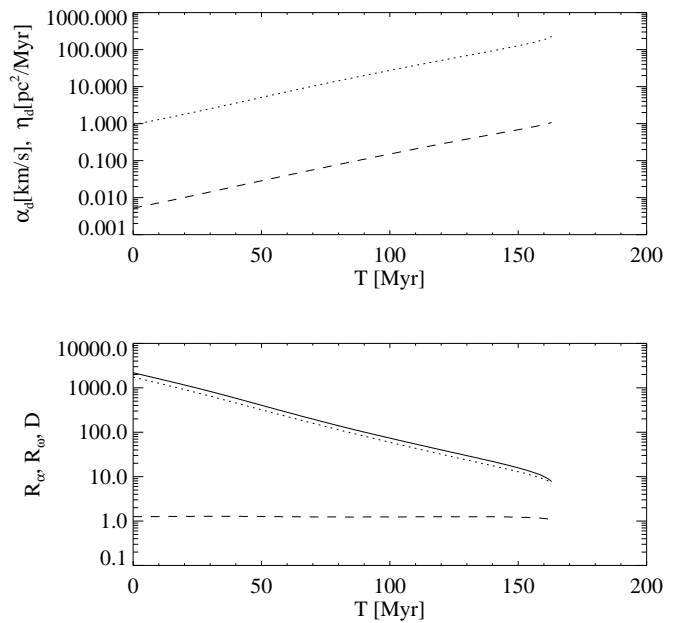}
\caption[]{
The same as in Fig.~5 in except of
the aerodynamic drag coefficient, which is $C_D=1$. This case corresponds to
Fig.~2. of this paper. }
\end{figure}
The dynamo coefficients based on our numerical simulations can be calculated
with the method applied in Paper I (formulae (82) and (84)).  As a first step
we shall present the temporal evolution of the helicity $\alpha_d$, diffusivity
$\eta_d$, The magnetic Reynolds numbers $R_\alpha$, $R_\omega$ and the dynamo
number $D$ for the parameter sets the same as that used in Figs.~1 and 2.  This
will allow us to make a first comparison of these coefficients resulting from
numerical simulations to that obtained in the linear approximation in Fig.~4 of
Paper I.

Fig.~5 representing the case without aerodynamic drag shows a close similarity
of the behaviour of the dynamo coefficients to the results obtained in the
frame of linear approximation. This similarity has been naturally expected
since the evolution of the shape is very similar in both the cases of
numerical simulations and the linear approximation. One minor difference is
apparent. Since the nonlinear growth is faster than the linear one for large
amplitudes of the Parker instability, the  growth of the transport
coefficients $\alpha_d$ and $\eta_d$ is faster in the nonlinear case as well.

The time dependence of the dynamo coefficients results from specific initial
conditions taken into considerations:  we begin with initial states of a small
amplitude as described in the Section 3.  We trace only the evolution of a
single unstable flux tube.  Since all the physical quantities involved are time
dependent, the dynamo coefficients are time dependent as well.  In more
elaborative approach averaging over a statistical ensemble of flux tubes with a
full spectrum of wavelengths and amplitudes of perturbations should be taken
into account.  This would allow to obtain saturated dynamo coefficients in
galactic discs without density waves.

The case of the aerodynamic drag coefficient $C_D=1$ is presented in Fig.~6.
We notice that the aerodynamic drag diminishes the dynamo coefficients
$\alpha_d$ and $\eta_d$ with respect to the case $C_D=0$, because the motion of
the flux tube is slowed down.

It is now worthwhile to check if one of the main conclusions of Paper I,
concerning the $\alpha$-effect in the weak magnetic field limit, is still valid
in the nonlinear range.  We shall examine the temporal evolution of the
$\alpha_d$ over the $(\alpha,\beta)$-plane.  The simulations have
been performed for a greed 10 times 5 of parameter points, covering the range
$0.0 \leq \beta \leq 1.0$ and $0.0 \leq \alpha \leq 0.5$.  The choice of the
wavelength maximizing the $\alpha_d$ coefficient has been made in the same way
as in Paper I.  All the simulations ended at the top singularity, while
the top of the flux tube has been displaced by the distance approximately equal
to the vertical scale length $H$.  At the same time the values of the
$\alpha_d$ were always close to 1, sometimes even larger.  The
results are shown in Fig.~7, where we plot $\alpha_d$ over the
$(\alpha,\beta)$-plane for $T = 40$ and $70 \Myr$.  In all the
simulations the initial vertical displacement amplitude is 10 pc.

\begin{figure}
\epsfxsize=\hsize \epsfbox{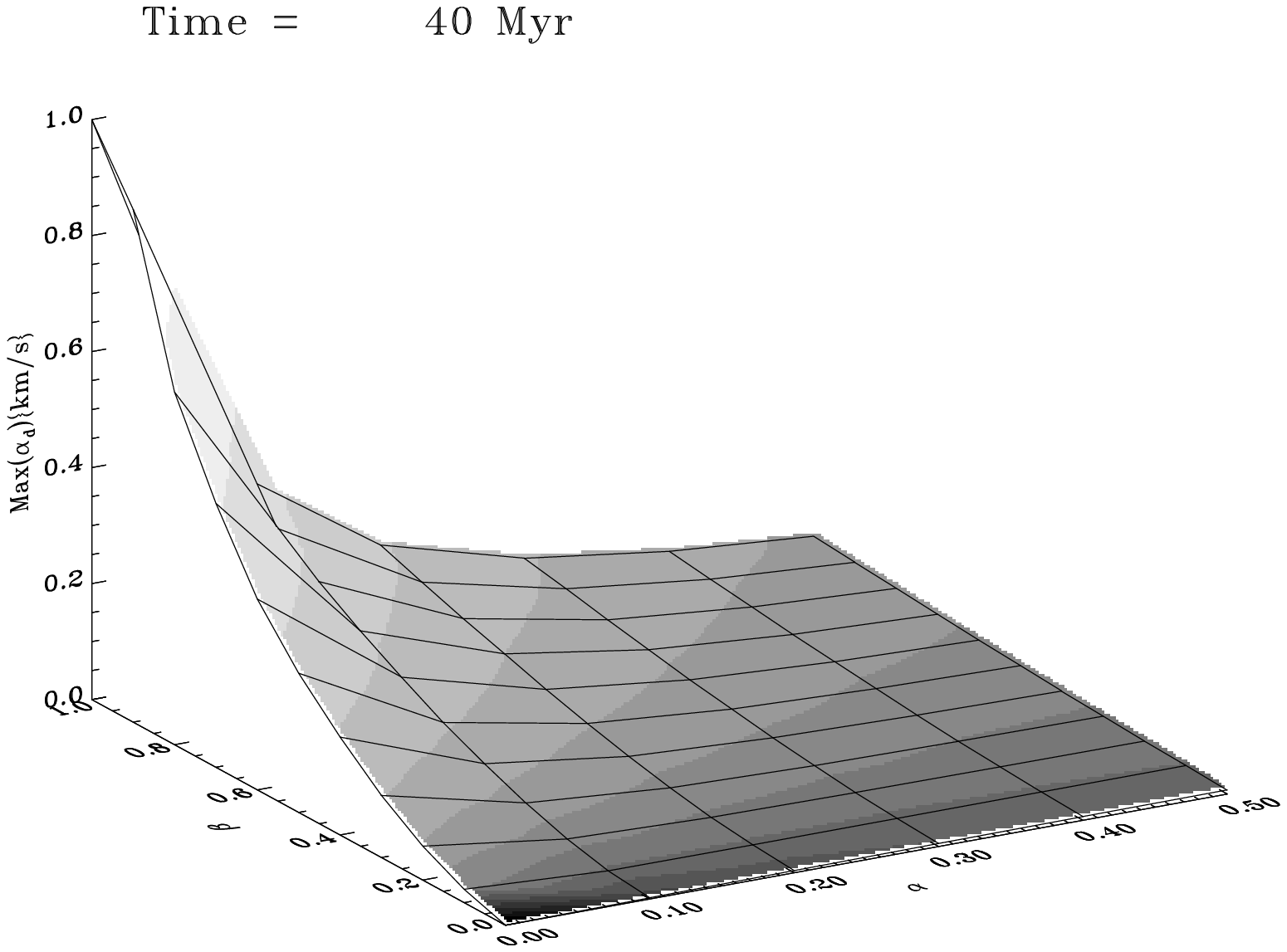}
\epsfxsize=\hsize \epsfbox{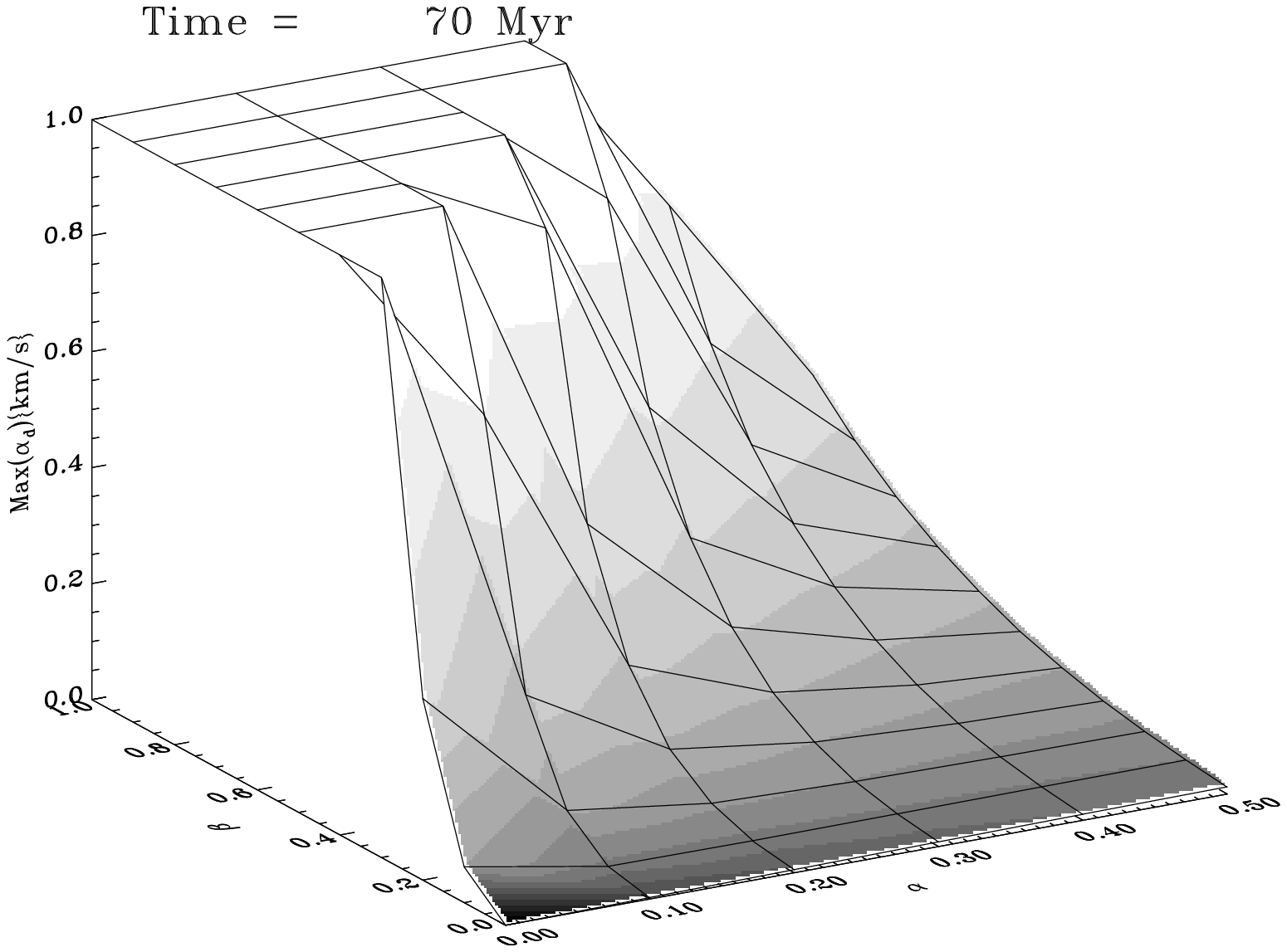}
\caption[]{
The temporal evolution of the $\alpha_d$ coefficient over the
$(\alpha,\beta)$-plane for $T=40 \Myr$ (upper panel) and $T = 70 \Myr$ (lower
panel). Because of the top singularity we  replaced the missing points with
the value $\alpha_d =1 \km/\s$ for $T=70 \Myr$.
}
\end{figure}

We note that the results are very similar to those obtained in frame of
the linear approximation.  We can conclude that the strongest dynamo
effect computed by means of the numerical simulations as well as in the linear
approximation is reached in a timescale of a few tens of Myr for weak magnetic
fields and the cosmic ray pressure comparable to the present value in
the Milky Way, $p_{cr} \sim 0.5\cdot 10^{-12} \dyn \cm^{-2}$.

\section{Nonlinear dynamics of magnetic flux tubes in a nonuniform external
medium}

We have shown in Paper I, that in the limit of weak magnetic field, the most
significant contribution to the helicity $\alpha_d$ is due to the short
wavelength modes.  But these modes are more stable than the long wavelength
modes which contribute to $\alpha_d$ only marginally (see Fig.~5 of Paper I).
This leads us to the question if the generation process of magnetic field can
still dominate over loses.  In a more detailed description the problem looks as
follows.  Let us assign

\begin{equation}
\lambda_{\alpha} \equiv \lambda (\max (\alpha_d)),
\end{equation}
for the wavelength maximizing the $\alpha_d$ coefficient in the linear
approximation, and $\lambda_{\max}$ for a certain very large wavelength for
which
the growth rate is close to maximum.  The respective growth rates for these
wavelengths are

\begin{eqnarray}
\omega_{i \alpha} \equiv \omega_i (\max (\alpha_d)), && \omega_{i \max} =
\max(\omega_i).
\end{eqnarray}
If we deal with a mixture of linear modes of various wavelengths which start
to grow with comparable initial amplitudes at $T = 0$, then the main
contribution to the total $\alpha_d$ comes from the mode of the
wavelength $\lambda_\alpha$.  In the linear approximation $\alpha_d$
grows with time proportionally to $\exp (2 \omega_{i \alpha} t)$. On
the other hand, the main contribution to the turbulent diffusivity comes from
the mode of the wavelength $\lambda_{\max}$ and grows proportionally to $\exp
(2\omega_{i \max} t)$ with $\omega_{i \max} > \omega_{i \alpha}$.  Then it is
possible that in the case of a spectrum of unstable modes, with a variety of
wavelength, the losses of magnetic field associated to the long wavelength
modes will become dominant over the generation process.

To check this hypothesis in the nonlinear range, for the flux tube
propagating in a uniform external medium we shall perform a numerical
experiment. We assume that the initial state is a composition of periodic
perturbations with wavenumbers given by

\begin{equation}
k_n = n k_0 = \frac{2 n \pi }{\lambda_0},
\end{equation}
where $\lambda_0$ is the maximum of wavelength determined by the
size of the computational domain, and $k_n$ are the successive wavenumbers
admitted by the periodic boundary conditions. The initial perturbations are
constructed in such a way that only the velocity is perturbed at $t=0$.
The initial displacements are identically equal to 0.

\begin{equation}
\bm{v}(s,t=0) =\sum_{n=1}^{n_{\max}}  \bm{v}_n \cos (k_n s + \varphi_n),
\end{equation}
where $n_{\max}=20$ is the number of periodic components and $\varphi_n$ is a
random phase given to each component for generality.  We assume that each
component has the same initial amplitude $v_n= 1 \km\s^{-1}$ and the initial
phases are generated by the random number generator.  We start simulations with
the above initial conditions and continue until the top singularity is reached.
Then we make a plot of the shape of the fluxtube at the time directly
preceding the top singularity.  An example is shown in Fig.~8.  We note that in
spite of a compact region of strong deformation due to the Parker instability,
the rest of the flux tube remains relatively straight and close to the
unperturbed equilibrium state.
\begin{figure}
\epsfxsize=\hsize \epsfbox{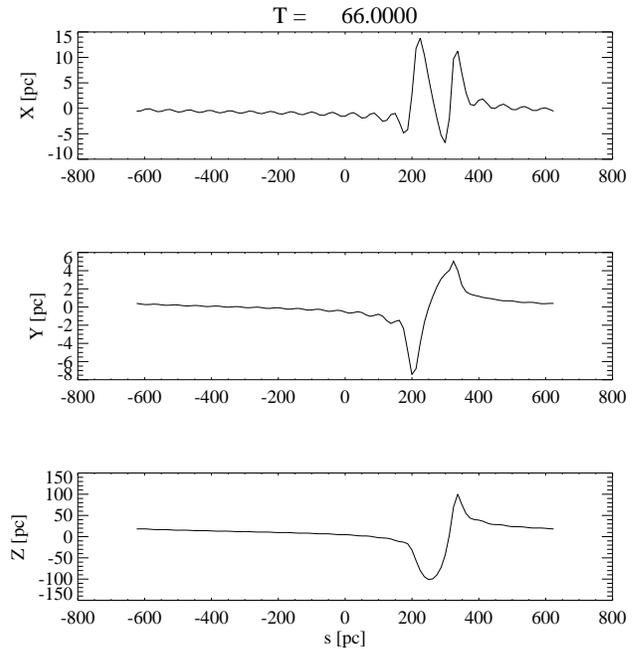}
\caption[]{The shape of the flux tube perturbed with a superposition of
periodic perturbations at the moment preceding appearance of the top
singularity at $T\approx 66 \Myr$
}
\end{figure}

\begin{figure}
\epsfxsize=\hsize \epsfbox{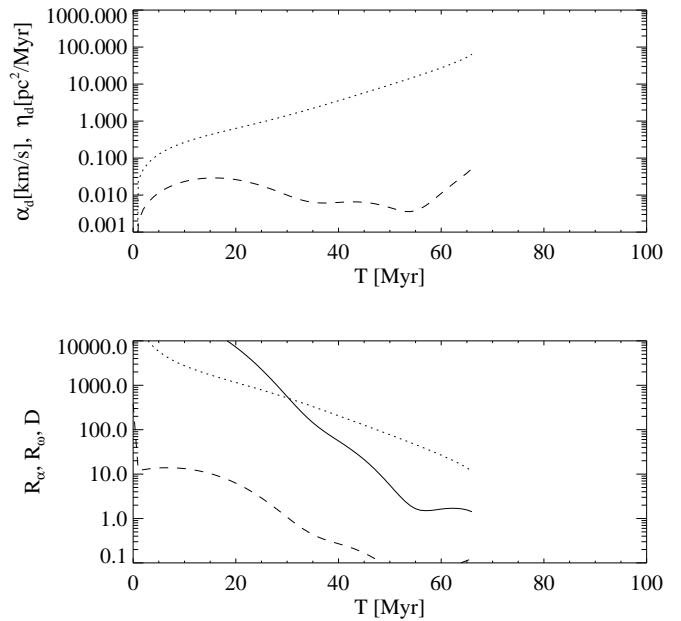}
\caption[]{The time dependence of the dynamo coefficients for the case
presented in Fig.~8.  Upper panel:  $\alpha_d$ (dashed line) and $\eta_d$
(dotted line).  Lower panel:  $R_{\alpha}$ (dashed line), $R_{\omega}$ (dotted
line) and $D$ (full line).  }
\end{figure}

The temporal evolution of the dynamo coefficients for this case is presented in
Fig.~9.  We easily note differences with respect to the highly idealized case
of the single component eigenmode initial perturbations.  According to our
expectations the dynamo coefficients $\alpha_d$ and $\eta_d$ are not
proportional to the same common exponential factor in the case of uniform
external medium.  The turbulent diffusion coefficient grows in an exponential
fashion apparently, while $\alpha_d$ varies in a chaotic way around a value
$\sim 0.01 \km \s^{-1}$, which is rather small.  For this reason
$R_{\alpha}$ is no longer close to a constant value, but diminishes
with time.  In consequence the dynamo number $D$ diminishes much faster than
$R_{\omega}$.  Then we note that the Parker instability would be very
destructive for galactic magnetic field if the case discussed above was
realistic.

The above case is not realistic however, because the interstellar medium is
very non-uniform.  The presence of molecular clouds is the apparent signature
of this fact.  We expect that the clouds perturb permanently the motion of flux
tubes.  If we assume, for instance, that the flux tubes are anchored to the
molecular clouds as proposed by Beck et al. (1991), (see HL'93, Paper I) then
the available wavelengths of the Parker instability are related to the mean
intercloud distance.  The longer wavelengths of the order of galactic disc
radius are suppressed due to the significant weight of molecular clouds and
the escape of magnetic field by means of the very long wavelength modes is
ineffective.

Alternatively, one can assume that the flux tubes are infinitely long and
uniform initially, but the ambient medium is cloudy.  In the forthcoming
considerations we shall apply a model of the flux tube -- external
medium interaction, resulting from the passage of the flux tube through a
medium of a nonuniform, variable density.  In order to limit a number of model
assumptions we do not take into account the aerodynamic drag force,
which in fact could very strongly influence the flux tube
motion.  We assume only that the density of external medium varies in a fashion
resembling a standing wave, whose maxima appear in variable places.
Let us propose an analytical model of such variations taking

\begin{equation}
\rho_e = \rho_{e0}(z) (1+
\frac{A_c}{2}(\sin(k_c s-0.5\omega_c t) -\sin(k_c s+1.5\omega_c t) )
\end{equation}
where the amplitude of the ``cloud wave'' $A_c = 0.2$, and the
wavelength $\lambda_c = 2\pi/k_c \sim 125 \pc$, equal to the most optimal
wavelength for the $\alpha$-effect (see Fig.~5 of Paper I), which on the other
hand is comparable to the typical intercloud distance 100 pc.  The frequency of
the external density modulations which we interpret as the
flux tube -- cloud collision frequency is

\begin{equation}
\omega_c = \frac{v_c}{\lambda_c}
\end{equation}
where $v_c$ is the typical molecular cloud speed $\approx 10 \km \s^{-1}$.  The
above form of the external density modulations was incorporated in the
subsequent simulation, where we applied the same initial conditions as
in the simulation leading to the results presented in Figs.~8 and 9.  The
simulation had been continued again until the top singularity was reached at $T
\sim 39 \Myr$.  The plot of the shape of fluxtube for this case is shown in
Fig.~10.
\begin{figure}
\epsfxsize=\hsize \epsfbox{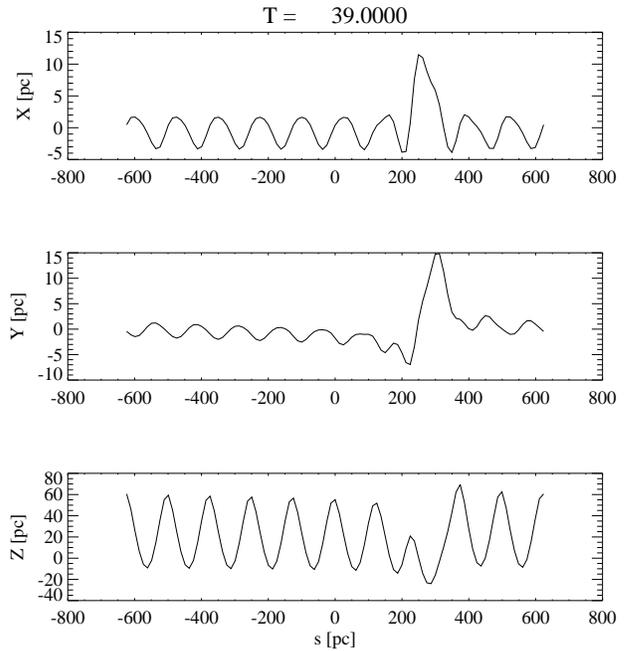}
\caption[]{The same as in Fig.~8, but with modulations of external density.
}
\end{figure}
\begin{figure}
\epsfxsize=\hsize \epsfbox{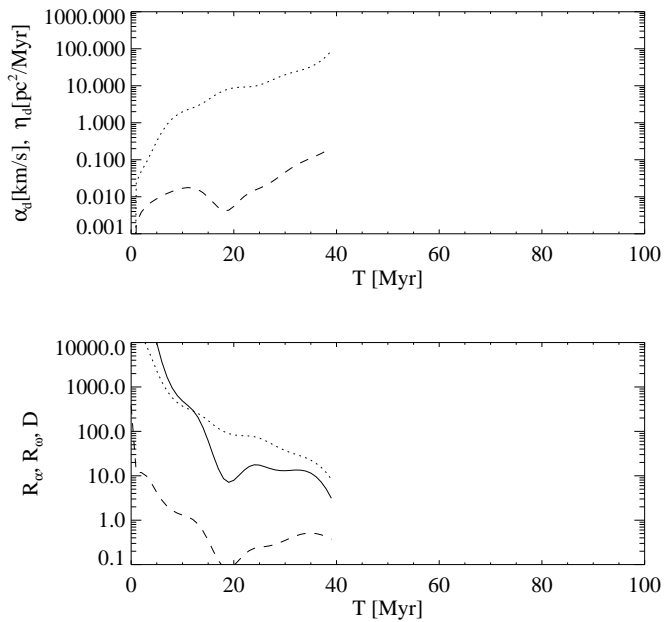}
\caption[]{The time dependence of the dynamo coefficients for the case
presented in Fig.~10.  Upper panel:  $\alpha_d$ (dashed line) and $\eta_d$
(dotted line).  Lower panel:  $R_{\alpha}$ (dashed line), $R_{\omega}$ (dotted
line) and $D$ (full line).  }
\end{figure}
We notice that in the presence of the external density modulations,
the whole length of the flux tube becomes buckled with the typical
wavelength equal to the wavelength of the external density modulations, what
is in contrast to the results obtained without the modulations.  The temporal
evolution of dynamo coefficients for this case is presented in
Fig.~11.

We note the essential differences with respect to the case without
the modulations of external medium:
\begin{enumerate}
\item The final value of the $\alpha_d$ coefficient is significantly
larger. Without external modulations we obtained $\alpha_d \sim 0.05 \km
\s^{-1}$, while
with modulations it is  $\sim 0.2 \km \s^{-1}$. The final value of the
$\eta_d$ coefficient is  approximately the same as in the case without
external modulations.
\item After an initial transient period of about 20 Myr, the transport
coefficients $\alpha_d$ and $\eta_d$ grow with a similar rate. This means that
modes of the same growth rate contribute to the both $\alpha$-effect and the
turbulent diffusivity.
\item The final value of the $R_\alpha$ coefficient is $\approx 0.5$ while
without modulations it was $\approx 0.1$. Let us recollect that $R_\alpha$
contains the ratio $\alpha_d/\eta_d$, hence the generation rate of radial
magnetic field is much more effective compared to losses in the
presence of external modulations.
\end{enumerate}
We can summarize then the role of the modulations of external medium as very
favorable for the dynamo action. This is because the long
wavelength modes, which are the most unstable in the
linear approximation, (we consider now the weak magnetic field limit) do not
dominate in the presence of the density modulations of external medium.
The $\alpha$-effect is strong due to the dominance of short wavelength modes.

\section{The effect of rotational shear}

In this section we  present the results of numerical simulations
taking into account  the differential force resulting from the rotational
shear ($F_{diff}= -4A\Omega\rho$).  As it has been described in Paper I the
magnitude of shear is parameterized by the Oort constant $A$, which is equal to
0, -1/2 and -3/4 for rigid, flat and Keplerian rotation curves
respectively.  The parameters of the simulations are:  $\alpha=\beta=1$,
$\Omega = \Omega_G$.  We fix the wavenumber $k_0= k_{crit} = 0.0044
\pc^{-1}$ ($\lambda_{crit} = 1430 \pc$) for $(A=-1/2)$.  Whereas the wavenumber
of perturbations is fixed, we shall vary the shearing parameter $A$ around
$A_0=-1/2$ corresponding to the flat rotation curve.  In Paper I we have shown
that the density waves introduce an additional differential force.
The differential force corresponding to the linearity limit of the interstellar
gas perturbations is equivalent to oscillations of $A$ with the 'top-bottom'
amplitude equal to about 50 \% of $|A_0|=1/2$.  This linearity limit
is related to a very small (1\%) spiral density wave perturbation of the
axisymmetric gravitational potential.

Let us take for simplicity two fixed values $A_1=-3/8$ and $A_2 = -5/8$ of the
Oort constant $A$.  The two values are placed symmetrically around $A_0$ and
their difference equals 50\% of $A_0$.  For $A_1$ the perturbation with $k_0$
is placed in the Parker range, while for $A_2$ in the shearing
range.  We shall perform two complementary runs of numerical simulations for
these two values of the Oort constant.  Our aim is to figure out the
qualitative differences between these two cases, considering the basic physical
variables already displayed in Figs.~3 and 4, and the dynamo coefficients, as
well.

Let us focus on the shape of the flux tube and the related variables first.
We shall analyze late stages of the evolution of the Parker-shearing
instability, which precede the appearance of the top singularity by less than 5
Myr.  The results are presented in Figs.~12 and 13.
\begin{figure}
\epsfxsize=\hsize \epsfbox{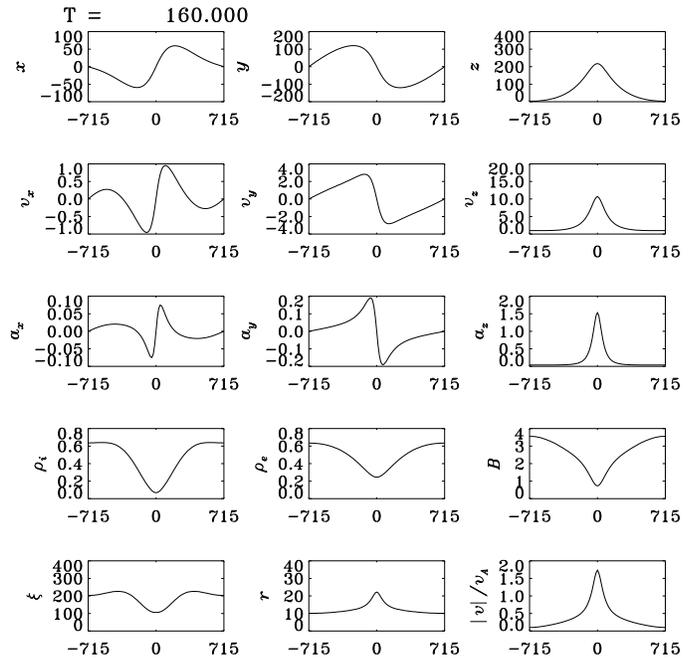}
\caption[]{The snapshot of the most relevant physical variables vs.  position
along the flux tube for the Parker range ($A = A_1 = -3/8$, $\lambda =
\lambda_{crit} = 1430 \pc$ for $A_0=1/2$).  The other parameters and units are
the same as in Fig.~3}
\end{figure}

\begin{figure}
\epsfxsize=\hsize \epsfbox{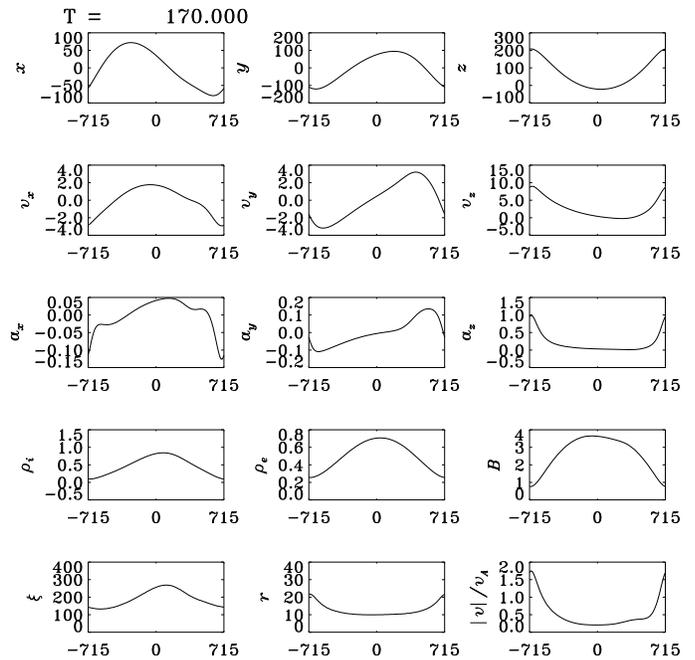}
\caption[]{The snapshot of the most relevant physical variables vs.  position
along the flux tube for the shearing range ($A = A_2 = -5/8$, $\lambda =
\lambda_{crit} = 1430 \pc$ for $A_0=1/2$).  The other parameters and units are
the same as in Fig.~3}
\end{figure}

In Fig.~12 ($A=A_1$, $k_0$ is in the Parker range) we note a regular,
symmetric shape of the flux tube with apparent signatures of some nonlinear
departures from the sinusoidal shape. The central part of the azimuthal
displacement $X$-curve is steepened, the top of $Z$-curve is vertically peaked
and
the bottom is shifted above the initial vertical position of the flux tube.  In
the linear regime the bottom displacement is negative and its absolute
value is the same as the vertical displacement of the top.  In contrast to the
case without shear, the azimuthal velocity curve $V_x$ becomes double
periodic, what means that gas does not flow toward the lowest point of the flux
tube at its lower parts, but instead it flows toward two points placed in
between the top and the bottom.  This is confirmed by the shape of the curve
representing the mass per unit length of the flux tube $\xi$.  We note two
points of the gas concentration in between the positions of the top and bottom.
The values of $\rho_i$ approaching 0 mean that the presented stage of the flux
tube evolution is close to the top singularity and the assumptions of the thin
flux tube approximation start to be violated near the top of the flux tube.
The $|v|/v_A$ exceeding 1 means that the transport of cosmic rays along the
magnetic field lines starts to be inefficient to inflate freely the topmost
parts of the flux tube.

In Fig.~13 ($A=A_2$, $k_0$ is in the shearing range) we note that the shape of
the flux tube becomes irregular.  We have demonstrated in Paper I that the
solutions are the propagating waves in this range, so that the azimuthal
position of the top is shifted with respect to its initial position at $s =0$.
The phase shift between the $y$- and $z$-displacements is less favorable for
the magnitude of the $\alpha_d$ coefficient than in the Parker range.  In the
linear range the formula for $\alpha_d$ contains the quantity $\Im (Y_1/Z_1)$
(the imaginary part of radial to vertical displacement amplitude ratio), which
is the largest for the phase difference about $\pm \pi/2$ between the vertical
and horizontal displacements.  Since this case represents the shearing range,
$\alpha_d$ is significantly reduced with respect to the Parker range as it is
shown in Fig.~8 of Paper I.  From the graph presenting the distribution of mass
per unit length $\xi$ we learn that gas concentrates around a single point,
which is maximally displaced in the positive $y$ (radial) direction.
Simultaneously this is the lowest point on the flux tube.  The point maximally
displaced in the negative $y$ direction has the lowest linear mass density and
is the highest point of the flux tube.  Similarly to the previously discussed
case of $A=A_1$, the presented stage of the flux tube evolution is at the limit
of applicability of the used approximations.

Comparing the above two cases we can say that the final mass distribution along
the flux tube is a result of compromise between two different redistributing
forces:  the gravitational force, which tends to place gas in the lowest point
of the flux tube and the differential force, which tends to place gas in the
points which are extremely displaced in the radial direction.  In the Parker
range (the differential force is relatively weak), the two mass concentrations
appear without a modification of the phase relations between the displacements
in three spatial directions.  For the stronger differential force in the
shearing range the phase relations are modified and the two categories of
concentration points merge (i.e.  the bottom becomes maximally displaced in the
radial direction).  This effect diminishes the dynamo $\alpha$-effect
forcing the disadvantageous phase relation between the radial and azimuthal
displacements.  It appears, however, that a residual $\alpha$-effect remains in
the shearing range, but it is much weaker than the $\alpha$-effect in the
Parker range.

The temporal evolution of the dynamo coefficients for both the cases of
Parker-shearing instability is presented in the subsequent Figs.~14 and 15,
which are complementary to Figs.~12 and 13 respectively.  We note that
$\alpha_d$ is apparently larger in the Parker range (Fig.~14) than in the
shearing range (Fig.~15), what confirms the previous results obtained in the
linear regime.  It is worth noting, however, that the dominance of the
$\alpha$-effect in the Parker range diminishes slightly during the nonlinear
evolution of the Parker-shearing instability.  The diffusivity $\eta_d$ behaves
very similarly in both the cases because the growth rates are very similar.
This implies the similarity of the magnetic Reynolds number $R_\omega$. The
magnetic Reynolds number $R_\alpha$ and in consequence the dynamo number $D$
are larger in the Parker range than in the shearing
range following the same behaviour of $\alpha_d$.

\begin{figure}
\epsfxsize=\hsize \epsfbox{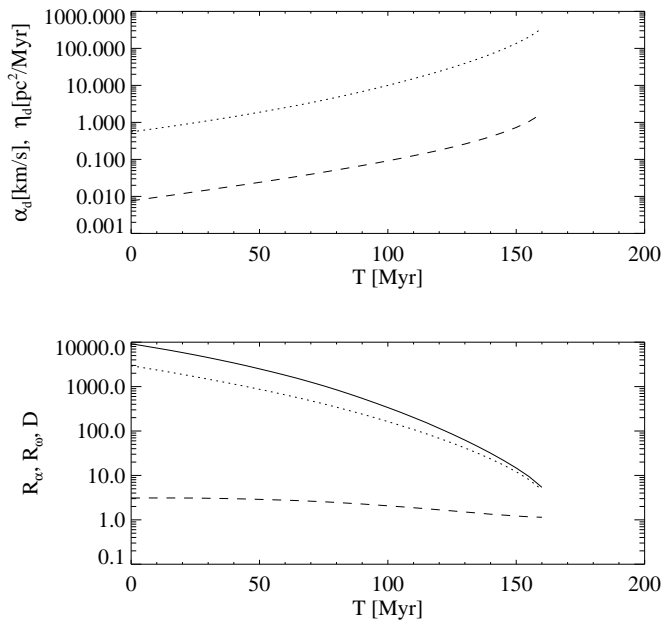}
\caption[]{The time dependence of the dynamo coefficients for the Parker range
($A=A_1$).  Upper panel:  $\alpha_d$ (dashed line) and $\eta_d$ (dotted line).
Lower panel:  $R_{\alpha}$ (dashed line), $R_{\omega}$ (dotted line) and $D$
(full line).  }
\end{figure}

\begin{figure}
\epsfxsize=\hsize \epsfbox{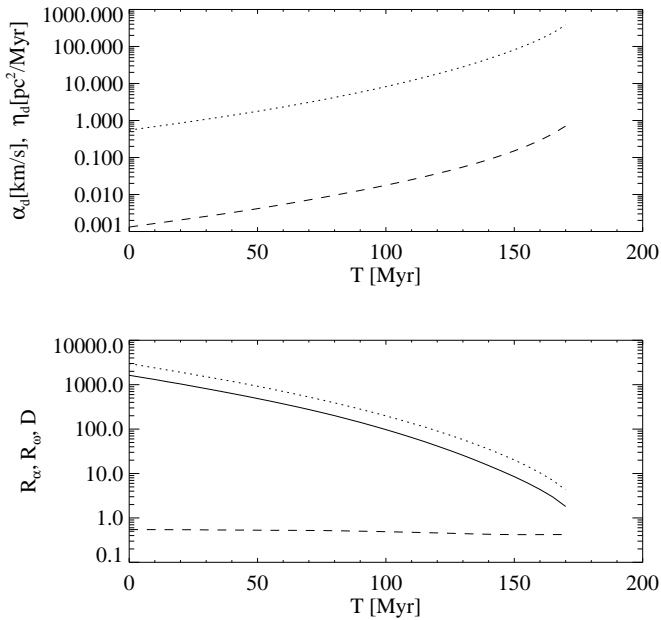}
\caption[]{The time dependence of the dynamo coefficients for the shearing
range ($A=A_2$).  Upper panel:  $\alpha_d$ (dashed line) and $\eta_d$ (dotted
line).  Lower panel:  $R_{\alpha}$ (dashed line), $R_{\omega}$ (dotted line)
and $D$ (full line).  }
\end{figure}

We can conclude that the nonlinear numerical simulations well confirm our
previous linear results of Paper I, with the exception that the difference
of $\alpha_d$ between the Parker and shearing ranges becomes less pronounced
in the nonlinear range.

\section{Discussion and Conclusions}

In this paper we present the results of numerical simulations solving the
nonlinear system of partial differential equations, derived in Paper I for the
dynamics of magnetic flux tubes in galactic discs.  The obtained numerical
solutions appeared to be very consistent with the linear analytical solutions
derived in Paper I for amplitudes of vertical displacements of the flux tubes
smaller than the vertical scale length $H$.  This consistency between the
numerical and the analytical results allows us to confirm the previous
conclusion: the weak magnetic field enables the fast growth of Parker
instability and the strong dynamo action due to the weakness of magnetic
tension if only the cosmic ray pressure is high enough.

An essential advantage of the performed simulations follows from the fact that
we have been able to trace the nonlinear evolution of the Parker unstable flux
tubes.  We have shown that due to the contribution of cosmic rays the top of
the flux tube inflates rapidly at finite vertical displacements of the
order of vertical scaleheight $H$.  This gives rise to an enormous growth
of the vertical buoyancy acceleration.  This process is able to form the
vertical structure of magnetic field lines extending to the galactic halo, as
it is observed in some cases of edge-on galaxies like NGC 4631 (Hummel et al.
1991; Golla and Hummel 1994) and M82 without any additional driving force.  We
note however that the inflation rate should be reduced if only the streaming
speed of cosmic rays along the magnetic field lines is limited, so that the
velocities of vertical rise should saturate at the level comparable to the
Alfven speed, typically several tens to hundreds of km/s.  The last estimation
depends on the particular values of the magnetic field strengths and gas
densities, so that large variations of the Alfven speed are possible depending
on galaxy.

Reuter et al.  (1992, 1994) point out that the halo of M82 is characterized by
a filamentary structure with prominent gaps localized between these filaments.
These features exhibit a preferential orientation normal to the plane of M82
and extend out to z-distances of $\simeq 1 \kpc$.  The linear polarization up
to 35 \% was found there with the polarization degree increasing with
z-distance.  Moreover, the vertical dependence of the spectral index of
synchrotron radiation was interpreted as an evidence, that the relativistic
plasma is streaming into the halo along vertical magnetic field lines with
velocities in excess of $1000 \km \s^{-1}$.  The M82 galaxy is known for
its intense star formation activity at the center, what implies the intense
production of cosmic rays, which are so important for our model.  We can say
that even if the Parker instability is not easily separable from the other
effects like galactic winds and galactic fountains, the presented observational
picture of the mentioned galaxies is consistent with our theoretical model.  We
would like to point out that we can explain the vertical magnetic field
structure of M82, the streaming of cosmic ray gas with very large vertical
velocity as well as the generation of magnetic field within a single model of
galactic dynamo driven by the Parker instability.

We have made a point in this paper that if the cosmic ray pressure
significantly exceeds the magnetic pressure, the Parker instability is able to
produce an effective dynamo action.  This is possible if 1) the flux tubes are
anchored in molecular clouds with mean separation comparable to the
most optimal wavelength for the $\alpha$-effect (see Paper I), or 2) the
flux tubes propagate in the nonuniform external medium as it has been
discussed in this paper.  Taking into account an exemplary model of weak
modulations of the external medium density we have shown that the dominating
wavelengths of the buckling of flux tubes is the same as the wavelength of
modulations of the external medium.  We attribute these modulations to the
presence of molecular clouds with typical separation of 100 pc.  With these
modulations the dynamo $\alpha$-effect is essentially stronger than without the
modulations and strong in comparison to the vertical diffusion of magnetic
field.  Without these modulations the diffusion rate tends to dominate over the
generation rate.

Performing numerical simulations of the nonlinear evolution of flux
tubes we have examined the role of the galactic differential forces resulting
from the axisymmetric differential rotation for two distinct values of the Oort
constant A. These simulations allowed us to figure out the qualitative
differences between the Parker and the shearing ranges introduced in Paper I.
We have found that for given perturbation with wavenumber $k_0$, in the Parker
range (weak shear) of the Parker-shearing instability, the qualitative
properties are similar to the case without shear. There is, however, an effect
resulting from the presence of the radial differential force -- the gas starts
to gather not at the lowest point of the flux tube, but at two points
extremely displaced in the radial direction. In contrast, we note a drastic
change of the general behaviour of the instability in the shearing range. The
perturbations become propagating waves. The gas starts to gather around one
point,
which is the lowest one and simultaneously the most displaced in the radial
direction. This qualitative change implies the modification of the phase
difference between the radial and the vertical displacements. This effect in
turn is responsible for the drop of the dynamo $\alpha_d$ coefficient.
In the nonlinear regime the $\alpha$-effect is still much more effective in
the Parker range than in the shearing range, but the jump of its magnitude is
smaller than that obtained in the linear regime.  Following Paper I we
associate the Parker range with the enhanced $\alpha$-effect with the spiral
arms.  Similarly we associate the shearing range with the diminished
$\alpha$-effect with the interarm regions.  We note that a more elaborated
approach should take into account the higher cosmic ray production rate in
spiral arms, which should additionally magnify the contrast of the
$\alpha$-effect between arms and the interarm regions.

Summarizing, we can say that all the numerical simulations well confirm
our former results of Paper I, and additionally highlight the new
nonlinear aspects.  In Paper I we proposed a new model of galactic dynamo
which is a cyclic process involving the action of the Parker-shearing
instability, galactic differential rotation, density waves and magnetic
reconnection.  In this model the $\alpha$-effect of the $\alpha\omega$ dynamo
is essentially dependent on the phase of the galactic density wave and is
localized in spiral arms.  The complementary shear (the
$\omega$-effect) is enhanced in the interarm regions. The dissipation
process necessary for the dissipation of the small scale nonuniform magnetic
field components is accomplished by the magnetic reconnection. Our dynamo model
is a kind of a fast dynamo  proposed by Parker (1992), which is free of
problems pointed out by Kulsrud and Anderson (1992).

Our approach, based on nonlinear dynamics of flux tubes, allowed us to compute
the dynamo transport coefficients with the back reaction of the large scale
magnetic field taken into account.  We found that the presence of cosmic rays
and the rotational differential force allow to overcome magnetic tension, so
that the dynamo transport coefficients do not saturate under the influence of
magnetic field of the strength below the observed equipartition value.  Thus,
our model avoids the basic problems of mean field dynamos as described by
Vainshtein and Cattaneo (1992).

\begin{acknowledgements}
I wish to thank prof. Harald Lesch and prof. Marek Urbanik for their
useful comments. This work was supported by the grant from Polish Committee
for Scientific Research (KBN), grant no. PB 0479/P3/94/07.
\end{acknowledgements}

\end{document}